\documentclass[twocolumn,showpacs,preprintnumbers,amsmath,amssymb]{revtex4}

\usepackage{graphicx}
\usepackage{dcolumn}
\usepackage{bm}

\begin{document}

\preprint{APS/123-QED}

\title{ Successive phase transition from superconducting to antiferromagnetic phase in (Ca$_6$(Al, Ti)$_4$O$_y$)Fe$_2$As$_2$ studied via $^{75}$As and $^{27}$Al NMR}

 \author{ T. Nakano$^1$,  N. Fujiwara $^1$ \footnote { Corresponding author: naoki@fujiwara.h.kyoto-u.ac.jp}, S. Tsutsumi$^1$, H. Ogino$^{2,3}$, K. Kishio$^{2,3}$ and J. Shimoyama$^{2,3}$
 }

\affiliation{$^1$ Graduate School of Human and Environmental
Studies, Kyoto University, Yoshida-Nihonmatsu-cyo, Sakyo-ku, Kyoto
606-8501, Japan}

\affiliation {$^2$ Department of Applied Chemistry, The University
of Tokyo, Tokyo 113-8656, Japan\ \\  $^3$ TRiP, Japan Science and
Technology Agency (JST), Sanban-cho bldg. 5, Sanban-cho, Chiyoda-ku,
Tokyo 102-0075, Japan}


\date{}


\begin{abstract}

An unusual successive phase transition from superconducting (SC) to
antiferromagnetic (AF) phases was discovered via $^{75}$As and
$^{27}$Al nuclear magnetic resonance (NMR) in
(Fe$_2$As$_2$)(Ca$_6$(Al, Ti)$_4$O$_y$) with four (Al, Ti)O layers
 intercalated between FeAs planes. Although the spatially-uniform AF
ordering is clearly visible from $^{27}$Al spectra, the ordered
moments are very small and the low-frequency fluctuation is much suppressed, contrary to existing pnictides with
localized magnetic elements. Furthermore, the temperature ($T$)
dependence of the fluctuation at both nuclei is very similar
throughout the entire temperature range. These facts suggest that
some hybridization between Ti and Fe orbitals induces
 a uniform electronic state within FeAs and (Al, Ti)O
 layers accompanied by the SC and AF transitions. The
 iron-based pnictide with Ti-doped blocking layers is the first high-$T_c$
compound having metallic blocking layers.

\end{abstract}

\pacs{74.70. Xa, 74.25. Dw, 74.25. nj, 76.60. -k}
\maketitle

To date, a variety of iron-based pnictides have been discovered
since the discovery of superconductivity in LaFeAsO$_{1-x}$F$_x$ (La
1111 series).$^1$ Among these compounds, the 1111 series
($R$FeAsO$_{1-x}$F$_x$ $R$=Ce, Pr, Sm, etc. ) and the 122 series
[Ba(Fe$_{1-x}$Co$_x$)$_2$As$_2$ ] have been extensively studied
because the highest superconducting transition temperature ($T_c$)
has been marked in the former family$^{2, 3}$ and a large single
crystal is available for the latter. The main difference between
them is the distance between FeAs planes: The 0.8739 nm distance of
the La 1111 series is greater than the 0.65 nm distance of the Ba
122 series.$^{1, 4, 5}$ Despite this $\sim$0.22 nm difference, their
electronic phase diagrams completely differ from each other. For the
1111 series, an overlap of the stripe-type antiferromagnetic (AF)
and superconducting (SC) phases is absent and the optimal $T_c$ is
realized away from the phase boundary, while for the 122 series, the
two phases overlap and the optimal $T_c$ is realized at the phase
boundary.$^{1-3, 6-8}$ The difference gives an impression that the
two families are entirely different. However, CaFe$_{1-x}$Co$_x$AsF
(Ca1111 series)$^{9}$ having an intermediate distance of 0.8593 nm
was recently found to have an intermediate overlap,$^{10}$
suggesting that the phase diagrams show continuous variation
depending on the distance between FeAs planes. The shrinkage of the
distance increases the overlap, although the optimal $T_c$ remains
approximately the same among these compounds.

The question is what occurs in compounds having a large distance
between FeAs planes. Actually, pnictides with perovskite-type
blocking layers have a large distance of 1.3 $-$ 2.5 nm.$^{11}$ A
high $T_c$ $\sim$40 K is marked in these compounds. The highest
$T_c$ in this family has been marked at 47 K for a titanium oxide
(Fe$_2$As$_2$)(Ca$_4$(Mg, Ti)$_3$O$_y$).$^{12}$ For other compounds,
such as (Fe$_2$As$_2$)(Ca$_5$(Sc, Ti)$_4$O$_y$) and
(Fe$_2$As$_2$)(Ca$_6$(Sc, Ti)$_5$O$_y$), $T_c$ exceeds 40 K.$^{11}$
In (Fe$_2$As$_2$)(Sr$_4$(Mg, Ti)$_2$O$_6$) and
(Fe$_2$As$_2$)(Sr$_4$V$_2$O$_6$), $T_c$ reaches 35 and 37 K,
respectively.$^{13, 14}$ This family is characterized by small
lattice constants, a narrow As-Fe-As bonding angle, and a large
distance between FeAs planes, which cause the lack of the hole
pocket corresponding to the $\alpha$ Fermi surface in undoped
systems as theoretically investigated in
(Fe$_2$As$_2$)(Ca$_4$Al$_2$O$_6$).$^{15, 16}$ Compounds such as
(Fe$_2$As$_2$)(Ca$_4$Al$_2$O$_6$) and
(Fe$_2$As$_2$)(Sr$_4$Sc$_2$O$_6$) are under the existing paradigm in
the meaning that the blocking layers are insulators like other
iron-based pnictides;$^{15-17}$ however Ti doping would induce new
Fermi surfaces of Ti-orbital origin.$^{18}$ In
(Fe$_2$As$_2$)(Sr$_4$V$_2$O$_6$), the hybridization between V and Fe
orbitals is possible, but strong correlations in V orbitals result
in Mott-Hubbard-type insulating blocking layers.$^{19, 20}$ The
insulating blocking layers with localized magnetic moments has been
suggested from the experiments using several techniques.$^{21}$

For this family, the focus is on whether transition metals in
blocking layers are involved in the formation of Fermi surfaces. An
understanding would shed new light on the interplay between
magnetism and superconductivity and high-$T_c$ mechanisms. We focus
on (Fe$_2$As$_2$)(Ca$_6$(Al, Ti)$_4$O$_y$) ($y \sim 12$) with four
(Al, Ti)O layers intercalated between FeAs planes, as shown in Fig.
1(a).$^{22}$ The series boasts of a 2.2 nm distance between FeAs
planes. Isomorphic compounds with two and three (Al, Ti)O layers,
(Fe$_2$As$_2$)(Ca$_{n+2}$(Al, Ti)$_n$O$_y$) ($n$ = 2, 3) have also
been synthesized; however,  they are of an inferior quality than the
four-layered samples.$^{22}$  The present compound is suitable for
nuclear magnetic resonance (NMR) measurements because
$^{75}$As($I=\frac{3}{2}$) and $^{27}$Al($I=\frac{5}{2}$) NMR can
provide information about both FeAs and (Al, Ti)O layers on a
microscopic level. We measured NMR using a conventional pulsed NMR
method for powder samples. The chemical formula expected from the
crystal structure is (Fe$_2$As$_2$)(Ca$_6$(Al$_{0.5}$
Ti$_{0.5}$)$_4$O$_{12}$) in which Ti is formally nonmagnetic and
therefore nonmagnetic blocking layers are expected.$^{22}$ A
superstructure corresponding to some Al ordering was not observed
from the x-ray analysis, implying that Al or Ti ions are randomly
distributed in the blocking layers. The samples contained FeAs
impurities, but the concentration was less than 2\% in weight. The
impurity phase was not observable in the $^{75}$As NMR measurements:
it is known that FeAs exhibits a magnetic ordering at 70 K,$^{23}$
but we could not find any anomaly around 70 K in the $^{75}$As NMR
measurements. Figure 1(b) shows the resistivity: The onset of $T_c$
is estimated to be 39 K and zero resistivity is at approximately 30
K, which is approximately the same as $T_c$ determined from the
detuning of an NMR tank circuit at 36.8 MHz, as shown in Fig. 1(c).

\begin{figure}
\includegraphics{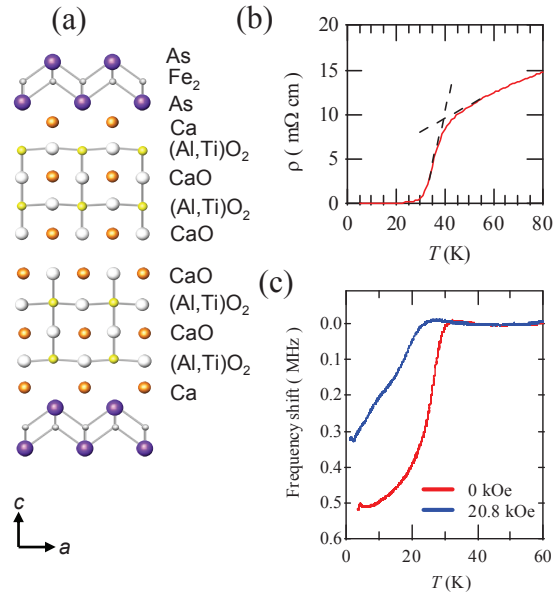}
\caption{\label{fig:epsart} (Color online) (a) Crystal structure of
(Fe$_2$As$_2$)(Ca$_6$(Al, Ti)$_4$O$_y$) [22]. (b) Temperature
dependence of the resistivity. The onset of $T_c$ is approximately
39 K [22]. (c) Deturning of a NMR coil measured at zero field and
20.8 kOe. The $T_c$ value at 20.8 kOe is estimated to be 23 $\pm$ 5
K. At this field, $^{27}$Al-$1/T_1T$ measurements were curried out,
as shown in Fig. 3(d).  }
\end{figure}

\begin{figure}
\includegraphics{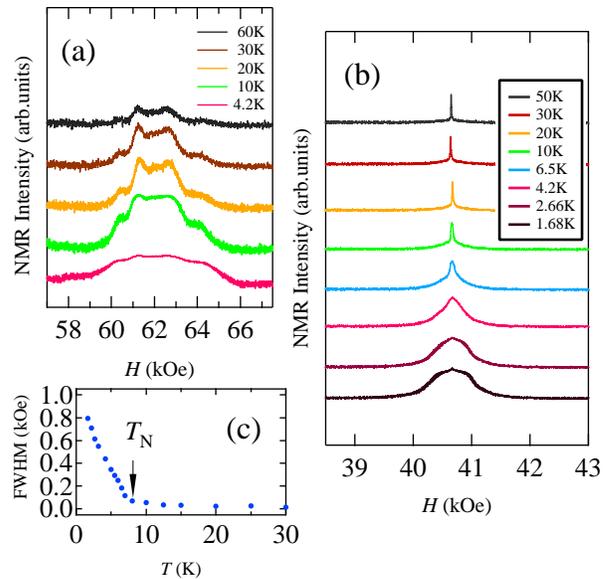}
\caption{\label{fig:epsart} (Color online) (a)$^{75}$As NMR spectra
at 45.1 MHz. The broadening at 4.2 K shows the appearance of some
spin ordering.
 (b) $^{27}$Al NMR spectra at 45.1 MHz. A rectangular-type powder pattern implies the appearance of spatially uniform AF moments. (c) The linewidth of $^{27}$Al NMR spectra. The ordering temperature is 7 K.  }
\end{figure}

\begin{figure*}
\includegraphics{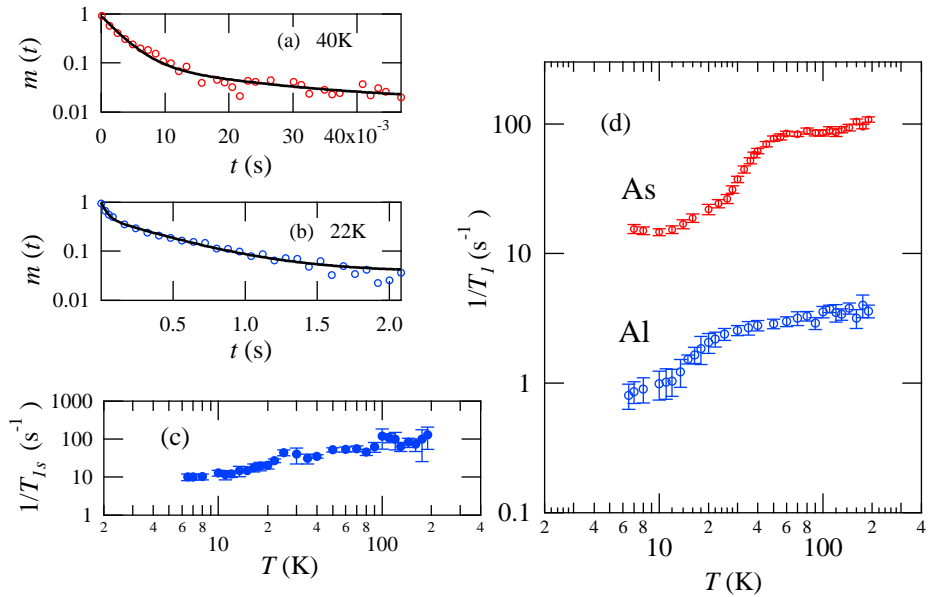}
\caption{\label{fig:epsart} (Color online) (a) Recovery curve,
$m(t)=1-M(t)/M(\infty)$ of $^{75}$As. Data are fitted by Eq. (1).
(b) Recovery curve of $^{27}$Al. Data are fitted by Eq. (2). (c)
Temperature dependence of the fast-recovery component $1/T_{1s}$
[See Eq. (2)]. (d) Temperature dependence of $1/T_{1}$ for $^{75}$As
and $^{27}$Al. }
\end{figure*}

Figures 2(a) and 2(b) show field-swept $^{75}$As and $^{27}$Al NMR
spectra, respectively, measured from spin-echo intensity at 45.1
MHz. The $^{75}$As NMR spectra for the central transition (I = -1/2
$\Leftrightarrow$ 1/2 ) broaden because of the second-order
quadrupole effect similar to other iron-based pnictides. At first
glance, the spectral pattern with four bumps is suggestive of two
$^{75}$As sites having different electric field gradients (EFGs). In
this case, however, the intensity difference between two $^{75}$As
sites is not explained. Instead, the lineshape is well reproduced by
considering a large anisotropic coefficient ($\eta $) of EFG. The
pure quadrupole frequency ($\nu_Q $) and $\eta $ are estimated to be
$\nu_Q $ = 13.4 MHz and $\eta \sim$ 0.6, respectively. The large
 $\eta $ value is rare in iron-based pnictides, although a fairly
large $\eta ( \sim$ 0.3 ) has been reported in
CaFe$_{1-x}$Co$_x$AsF.$^{10}$ Two bumps around 61 and 63 kOe
correspond to $\theta$ = 90$^{\circ}$ and 42$^{\circ}$,
respectively, where $\theta$ represents the angle between the $c$
axis and the applied field. The lineshape changes at low
temperatures reflecting some spin ordering. However, the position of
the low-field bump is unchanged even in the ordered state, implying
that the ordered moments are aligned parallel to FeAs planes like
other pnictides: For the samples associated with the low-field bump,
the stripe-type ordered moments result in the internal field
perpendicular to the applied field, and have no affect on the
low-field bump. The appearance of some ordering is clearly visible
from the $^{27}$Al NMR spectra. Although satellite signals (I = $\pm
\frac{3}{2} \Leftrightarrow \pm \frac{5}{2}$) are expected to appear
because of the quadrupole effect, only a sharp signal is observable
at high temperatures, implying that Al nuclei experience an EFG that
is very small and is somewhat distributed. The sharp signal also
indicates that the local environment around $^{27}$Al is hardly
affected by random Al distribution. The sharp signal changes to a
rectangular-type powder pattern at low temperatures. The ordering
temperature ($T_N$) is estimated to be 7 $-$ 8 K from the
temperature ($T$) dependence of the linewidth, as shown in Fig.
2(c). The tail of the rectangular pattern on both sides is caused by
the effect of small satellite signals. The rectangular-type powder
pattern is symmetric with regard to the free position corresponding
to the Larmor frequency, demonstrating that an AF spin configuration
is formed and the amplitude of spin moments is spatially uniform.

\begin{figure*}
\includegraphics{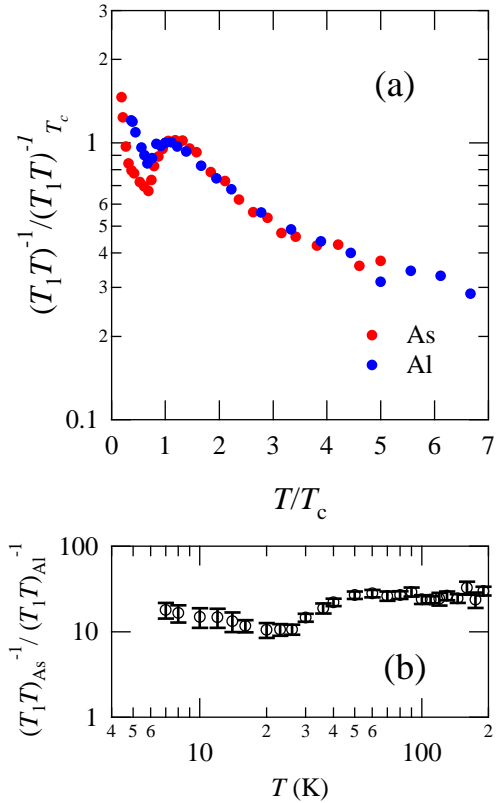}
\caption{\label{fig:epsart} (Color online)(a) Plot of $1/T_1T$
versus $T$ normalized by those at $T_c$. (b)The ratio of
$(1/T_1T)_{As}/ (1/T_1T)_{Al}$. The values are almost the same
except for the temperature range between 20 and 40 K corresponding
to $T_c$ values at zero field and 20.8 kOe.}
\end{figure*}

 Relaxation rates ($1/T_1$) of $^{75}$As and $^{27}$Al were measured using a saturation-recovery method.
 We measured $1/T_1T$ of $^{75}$As ( $(1/T_1T)_{As}$ ) at the lower-field bump and $1/T_1T$ of $^{27}$Al ( $(1/T_1T)_{Al}$ ) at a low field of 20.8 kOe because a large field could potentially break down the superconductivity or reduce the SC volume fraction.
  Prior to the measurements, we confirmed whether the superconductivity is maintained at 20.8 kOe from the detuning [Fig. 1(c)].
  At this field, $T_c$ is estimated to be 23 $\pm$5 K. The decrease in $T_c$ was also confirmed from the AC and DC susceptibility.$^{24}$ A
  remarkable decrease in $T_c$ under a low field may be a peculiarity of this system. Such caution is not required for $^{75}$As, because FeAs planes at
the low-field bump are parallel to the applied field, and the
decrease in $T_c$ under the field is neglected.
  Figures 3(a) and 3(b) show recovery curves, time variations of spin-echo intensity after a saturation pulse, for $^{75}$As and $^{27}$Al, respectively. Data of $^{75}$As follow the conventional rate equation for the
central transition (I = -1/2 $\Leftrightarrow$ 1/2 ),
\begin{equation}
 1-\frac{M(t)}{M(\infty)}=0.9exp(-6t/T_1)+0.1exp(-t/T_1),
\end{equation}
 while all $^{27}$Al satellite signals are saturated owing to the small quadrupole splitting, and data of $^{27}$Al exhibit a recovery with two components,
\begin{equation}
1-\frac{M(t)}{M(\infty)}= (1-c) exp(-t/T_{1s})+ c exp(-t/T_1).
\end{equation}
The $T$ dependence of the fast- and slow-recovery components is
shown in Figs. 3(c) and 3(d), respectively, together with
$(1/T_1T)_{As}$. The value of c increases approximately linearly
from 0.4 to 0.6 with increasing temperature. The fast-recovery
component $1/T_{1s}$ hardly reflects the electronic state, and
possibly originates from the nuclear spin diffusion. Bending points
in Fig. 3(d) are in good agreement with $T_c$ determined from the
detuning in Fig. 1(c). $1/T_1$ values of $^{75}$As are one order
greater than those of $^{27}$Al, while the $T$ dependence, excluding
$T_c$ values, of both nuclei is very similar. The $T$ dependence of
$(1/T_1T)_{As}$ and $(1/T_1T)_{Al}$ normalized by the values at
$T_c$ is shown in Fig. 4(a). They both show Curie-Weiss behavior and
are almost the same at high temperatures above $T_c$, implying that
both nuclei experience the same spin fluctuation.

Interestingly, Curie-Weiss behavior returns in $1/T_1T$ at low
temperatures,  reflecting the AF ordering. The question is which of
Ti and Fe is responsible for the AF ordering. In the existing
paradigm, the blocking layers are insulators and the AF ordering
would occur either in FeAs or (Al, Ti)O layers. In both cases, some
of the results are not well explained.

(1)In the case of Fe origin, Fe orbitals are responsible for both SC
and AF orderings, which would be very exotic if possible.
  The possibility is ruled out because two inner (Al, Ti)O layers are away from FeAs planes and only weak dipole coupling is effective as hyperfine coupling,
  making the linewidth narrow and the spectral weight strong at the central position of the $^{27}$Al
  lineshape. The linewidth of $^{27}$Al in the inner (Al, Ti)O
  layers is estimated as $\sim$5 Oe when a Bohr magnetron is assumed as the amplitude of Fe
  moments.

(2) In the case of localized Ti moments, the appearance of the
spatially uniform AF spin moments results in a $3d^{1}$ state having
an electron at each Ti site, despite that Ti ions are expected to be
nonmagnetic from the crystal structure. Such case may be possible
when electrons are transferred from FeAs planes to (Al, Ti)O layers.
If all Ti ions were magnetic, Al nuclei would experience strong
low-frequency fluctuation, considering the location of Al nuclei.
This is seen in other compounds such as SmFeAsO$_{1-x}$F$_x$ and
NdFeAsO$_{1-x}$F$_x$: Predominant spin fluctuation, namely
Curie-Weiss behavior in $1/T_1$, masks information from FeAs
planes.$^{25, 26}$ On the contrary,  $1/T_1$ is almost constant at
$^{27}$Al, and $T_c$ is observable at high temperatures in the
present compound.

 Experimental results are not perfectly explained by the existing
paradigm. The reason why (Al, Ti)O layers are not insulators is
derived directly from the amplitude of the ordered moments $<S_i>$.
The amplitude is estimated from the hyperfine coupling $^{27}A_{hf}$
and internal field $\Delta H$ of $^{27}$Al as $ \Delta H = g\mu_B
<S_i>^{27}A_{hf}$. The coupling $^{27}A_{hf}$ mainly originates from
the Fermi contact rather than the super-transferred hyperfine
coupling. Because the coupling is rather isotropic, all fluctuations
are observable at $^{27}$Al sites, but in-plane AF fluctuations
would be predominant as investigated in the 122 series.$^{27}$ In
the case of $^{75}$As, the hyperfine coupling $^{75}A_{hf}$ mainly
originates from the transferred hyperfine coupling. As nuclei
experience the in-plane AF fluctuations remarkably when the field is
applied parallel to FeAs planes.$^{27}$  Therefore, the in-plane AF
fluctuations would be predominant at both $^{27}$Al and $^{75}$As
sites. The coupling $^{27}A_{hf}$ can be estimated by the following
relation assuming that both $^{27}$Al and $^{75}$As experience the
same AF fluctuation:

\begin{equation}
 \frac{1}{(T_1)_{As}}/ \frac{1}{(T_1)_{Al}} \sim (\frac{^{75}\gamma_N}{^{27}\gamma_N})^2
(\frac{^{75}A_{hf}}{^{27}A_{hf}})^2.
\end{equation} The value is approximately 30 at high temperatures, as seen in Fig. 4(b). The gyromagnetic
ratios of $^{75}$As and $^{27}$Al, $^{75}\gamma_N$ and
$^{27}\gamma_N$ are 7.29 and 11.09 MHz/10 kOe, respectively. The
coupling $^{75}A_{hf}$ has been estimated to be 26 kOe/$\mu_B$ for
the field parallel to the FeAs planes,$^{27}$ and $^{27}A_{hf}$ is
estimated to be 3.1 kOe/$\mu_B$ from Eq. (3). The internal field $
\Delta H$ is $\sim$ 0.3 $-$ 0.4 kOe as seen in Fig. 2(b), therefore
$<S_i>$ is estimated to be $\sim 0.05 - 0.07$. The value is very
small compared with that expected from localized spin moments $<S_i>
= 1/2$. The small AF magnetization is a feature of weakly
antiferromagnetic metals. To explain the small amplitude of $<S_i>$
and the similarity between $^{75}$As and $^{27}$Al sites in $1/T_1T$
[Fig. 4(b)],$^{28}$ some hybridization between Ti and Fe orbitals
would be crucial.

 For (Fe$_2$As$_2$)(Sr$_4$V$_2$O$_6$) with a high $T_N$ of
150 K,$^{29}$ V orbitals should be removed from the Fermi level
owing to strong correlations of $\emph{d}$ orbitals in vanadium;
namely a Mott-Hubbard-type insulating state is realized. Taking
account of a small $T_N$ ( $\sim$ 7 K) caused by Al-rich doping,
correlations are expected to be small, and therefore, (Al, Ti)O
layers are responsible for Fermi surfaces, contrary to the VO
blocking layers. The appearance of the AF state following the SC
state implies that some parts of the Fermi surfaces are not involved
in the formation of a SC gap. In this sense, the phenomenon is
similar to homogeneous coexistence of the incommensurate spin
density wave and superconductivity in the crossover regime of the
122 or 1111 series.$^{30-32, 10}$

In summary, an unusual successive phase transition from the SC to AF
phases was observed via $1/T_1T$. Although the spatially uniform AF
ordering is clearly visible from $^{27}$Al spectra, localized spin
moments are unlikely to exist in (Al, Ti)O blocking layers. The
ordered moments are very small, and the $T$ dependence of $1/T_1T$
at both $^{75}$As and $^{27}$Al sites is very similar throughout the
entire temperature range. These facts suggest that some
hybridization between Ti and Fe orbitals induces a uniform
electronic state within FeAs and (Al, Ti)O layers accompanied by the
SC and weak AF transitions.

The authors would like to thank K. Kuroki for discussion. The
present work was partially supported by a Grant-in-Aid (Grant No.
KAKENHI 23340101) from the Ministry of Education, Science, and
Culture, Japan.







\end{document}